\documentclass[a4paper,11pt]{article}
\usepackage{pos}
\usepackage{hyperref}
\usepackage{aas_macros}
\usepackage{wrapfig}
\newcommand{\C}{3C\,84}

\begin{document}
\title{Pinpoint the jet apex in \C}
\ShortTitle{\C\ jet apex}
\author*[a]{G.~F.~Paraschos}
\author[b,a]{J.-Y.~Kim}
\author[a]{T.~P.~Krichbaum}
\author[c]{J.~Oh}
\author[c, b]{J.~A.~Hodgson}
\author[d]{M.~A.~Gurwell}
\author[a]{J.~A.~Zensus}
\affiliation[a]{Max-Planck-Institut für Radioastronomie, Auf dem Hügel 69, 53121, Bonn, Germany;}
\affiliation[b]{Korea Astronomy and Space Science Institute, 776 Daedeokdae-ro, Yuseong-gu, Daejeon 30455, Korea}
%\affiliation[c]{Department of Physics and Astronomy, Seoul National University, 1 Gwanak-ro, Gwanak-gu, Seoul 08826, Korea}
\affiliation[c]{Department of Physics and Astronomy, Sejong University, 209 Neungdong-ro, Gwangjin-gu, Seoul 05006, Korea}
\affiliation[d]{Center for Astrophysics \textbar ~Harvard \& Smithsonian, 60 Garden Street, Cambridge, MA 02138, USA}

\emailAdd{gfparaschos@mpifr-bonn.mpg.de}

\abstract{Jets which are powered by an AGN are a crucial element in the study of their central black holes (BH) and their immediate surroundings. 
The formation of such jets is the subject of intense research, mainly based on the dichotomy presented by the two main jet launching scenarios -- the one from Blandford \& Payne (1982), and the one from Blandford \& Znajek (1977).
In this work we study the prominent and nearby radio galaxy \C\ (NGC\,1275) with 15, 43, and 86\,GHz quasi-simultaneous VLBI observations. From these we determine the jet apex to be located $83\pm7$\,$\mu$as ($0.028-0.11$\,pc) upstream of the 86\,GHz VLBI core, applying a two dimensional cross-correlation analysis.
A byproduct of this analysis are spectral index maps, in which we identify a robust spectral index gradient in the north-south direction, for the first time at such high resolution, for the 43-86\,GHz pair.
The magnetic field strength at distances from the VLBI core comparable to measurements from the literature ($\sim10$ Schwarzschild radii) for other prominent AGN, like NGC\,1052 and M\,87, is computed to be $70-600$\,G.
Implications for the magnetic field topology are also discussed.
}

\FullConference{%
  *** European VLBI Network Mini-Symposium and Users' Meeting (EVN2021) ***\\
  *** 12-14 July, 2021 ***\\
  *** Online ***
}
\maketitle

\section{Introduction -- \C\ over the years}
\begin{figure}
    \centering
    \includegraphics[width=1\columnwidth]{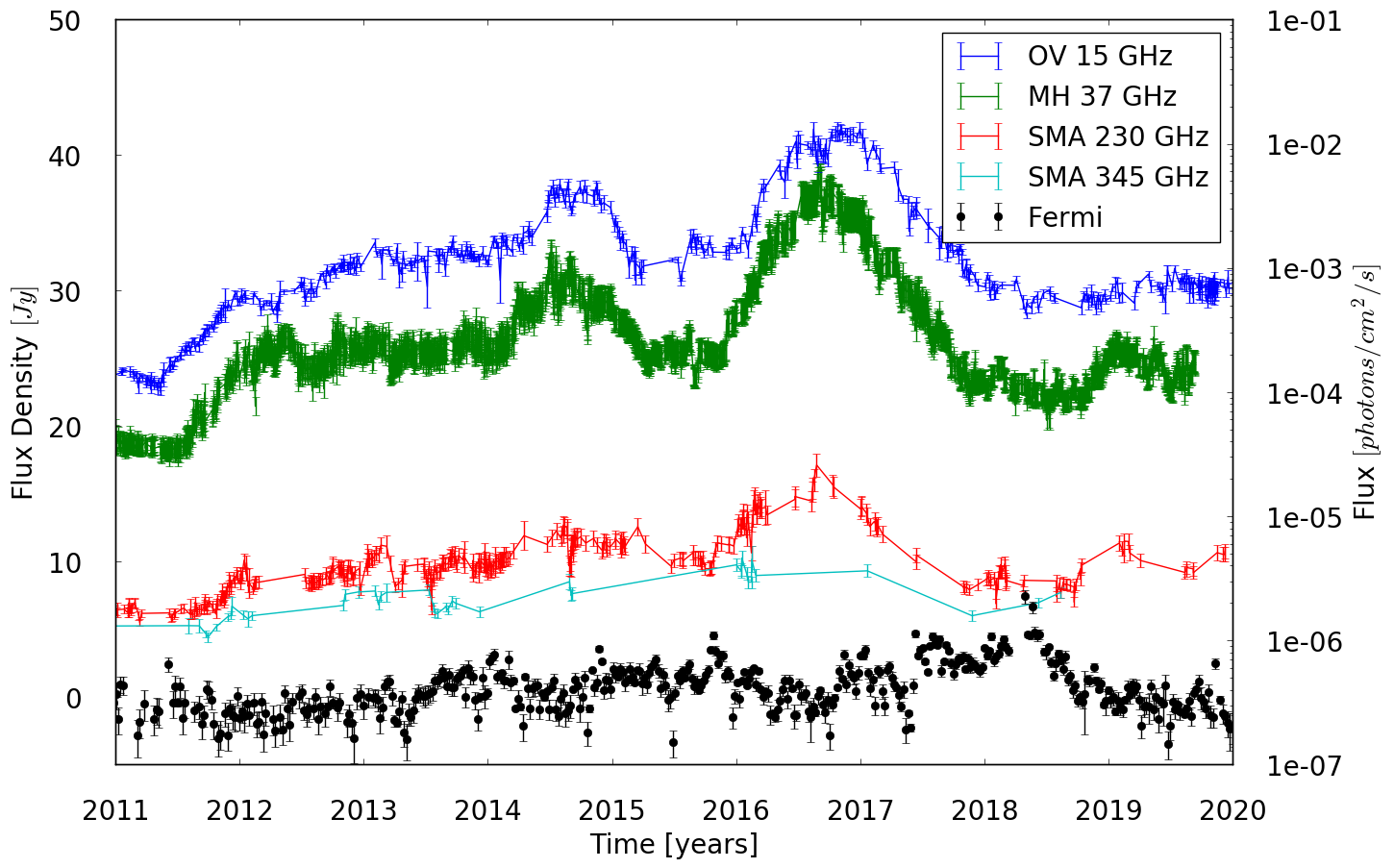}
    \caption{\small \sl Radio flux of \C\ at 15\,GHz (Owen's Valley Radio Observatory, blue; see \citep{Richards11}), 37\,GHz (Metsähovi Radio Observatory, green), 230\,GHz, and 345\,GHz (Submillimeter Array, red and teal respectively), as well as the $\gamma$-ray flux (Fermi-LAT, black; see \cite{Kocevski21}) since the year 2011.}
    \label{fig:flux}
\end{figure}

\noindent
NGC\,1275 is the central radio galaxy of the Perseus cluster. 
It harbours \C, which is one of the brightest radio galaxies in the northern sky, making it a suitable target for high spatial resolution imaging with centimetre- and millimetre-VLBI. 
From radio to $\gamma$-rays, the source shows prominent flux variability (see Fig. \ref{fig:flux}). Its relation to the structural variability detected in the VLBI images is a topic of intense study.
The first VLBI images of \C\ were published in the early 1950s \citep{Baade54}. Early cm- \cite{Walker00} and mm-VLBI maps \cite{Dhawan90, Krichbaum92} already showcased the complexity of the jet both near the nucleus, as well as further downstream, in the $\sim 10-20$\,mas region, where a complex and lobe-like structure is formed by the more extended emission. At 43\,GHz and on the 3\,mas scale three
dominant emission regions, denoted with C1, C2, and C3 exist \cite{Nagai14}.
C1 is the nuclear region, C2 is the diffuse emission in the southwest of the nucleus and C3, which is a recently ejected component ($\sim$2005 \citep{Hodgson21}), travelling in a southern direction. Two major flux brightening events are known for \C; one in the 1990's, which might be connected to the appearance of C2 and one in the 1960's \citep{Dent66} from which the $10-20$\,mas diffuse emission in the south may originate.

The pronounced source activity likely
has resulted in the two sided jets we are observing nowadays, with the southern jets being more prominent.
The counter-jet is fainter and has been observed at 8, 22, 43 \citep{Vermeulen94, Walker94, Fujita17}, and recently also at 86\,GHz \cite{Wajima20}.
An optical thick free-free absorber (perhaps a torus) around the central engine has been proposed as a possible explanation for the lack of observable counter-jet emission on sub-mas scales \citep{Vermeulen94, Walker94, Kim19}.

Regarding the jet kinematics, components seem to advance in a dense medium at sub-luminal velocities ($\sim$0.1c) and then accelerate further downstream to $0.3-0.5$c, with reported apparent velocities of up to 0.9c \citep{Krichbaum92, Vermeulen94, Hodgson21}. The components seem to follow curved trajectories \citep{Krichbaum92, Dhawan98}.

Flux peaks can usually be associated with component ejections and intense activity in the jet region.
Figure \ref{fig:flux} displays the radio and sub-mm flux density, together with $\gamma$-ray emission, of \C\ since 2011.
The tight correspondence between the radio and sub-mm flux density is noticeable, with the intensity peaks in 2015 and 2017 apparently correlated with component ejections in the 0.5\,mas region south of the VLBI core, seen both at 43, and 86\,GHz observations (see Paraschos et al. in prep.).
% The flux depression seen in 2016 (at cm-wavelengths) could be associated with the observed frustration of the C3 component; the decrease in velocity might result in less pronounced Doppler boosting.
% In \cite{Kino21}, the frustration is explained by the collision between the jet and a compact, dense cloud. 

\C\ is also one of the very few TEV $\gamma$-ray emitting sources \citep{Britzen19, Linhoff21, Hodgson21}.
However, possible correlations between the $\gamma$-ray and radio flaring lack high significance (see Fig. \ref{fig:flux}).

\section{Jet apex position}
\begin{figure}
    \centering
    \includegraphics[width=1\columnwidth]{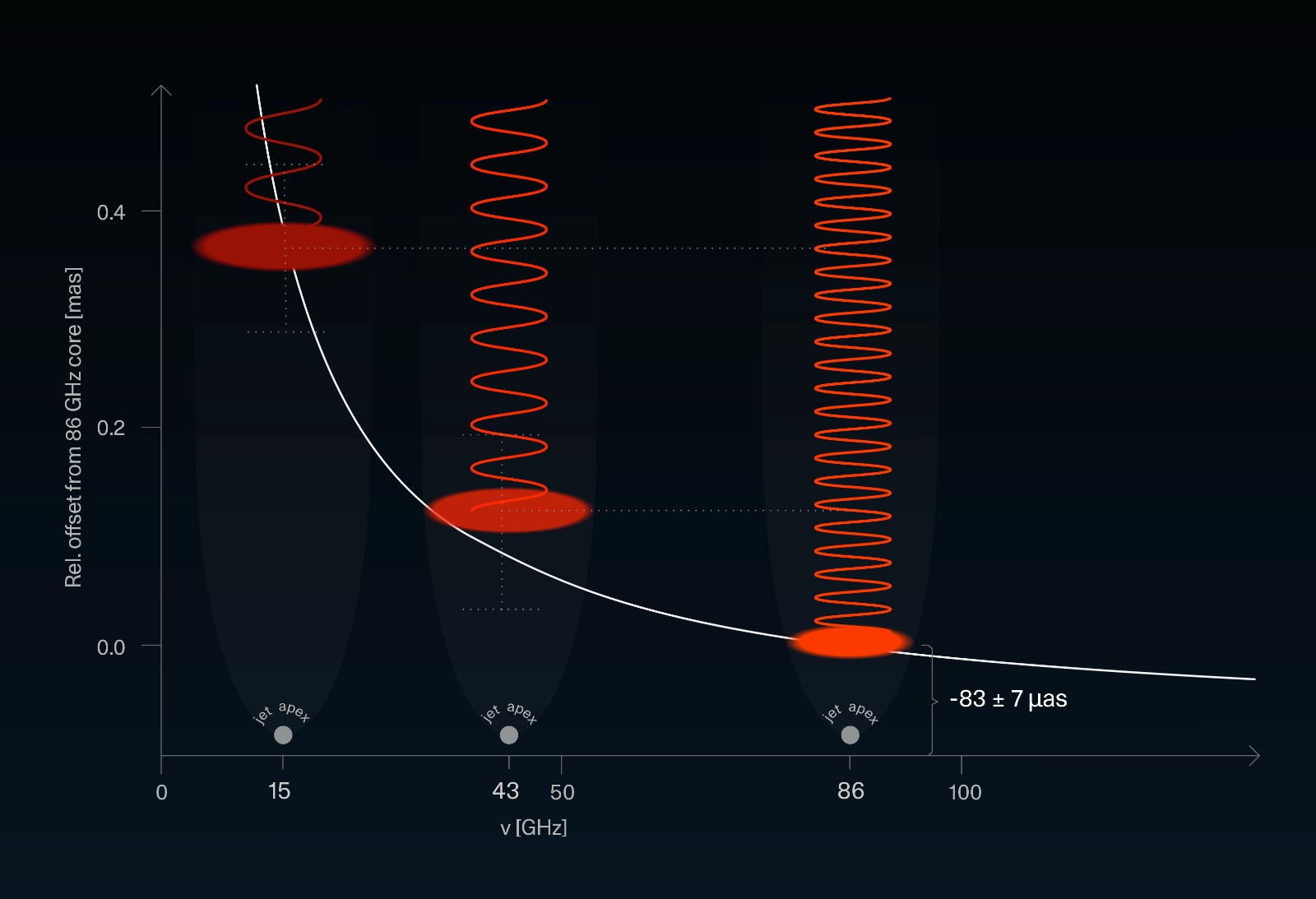}
    \caption{\small \sl Position offset of the 15 and 43\,GHz intensity peaks in the core region relative to the 86\,GHz intensity peak, as a function of frequency. The grey continuous line displays the fit of Eq. \ref{Eq:CS} to the data points. The distance of the jet apex (denoted with grey disks) to the 86\,GHz intensity peak is found to be $83\pm7$\,$\mu$as. The red/orange ellipses showcase the observed core at different wavelengths/frequencies, illustrating the concept of the core shift.}
    \label{fig:CS}
\end{figure}

\noindent
Quasi-simultaneous observations of \C\ were carried out at 15, 43, and 86\,GHz in May 2015, with the Very Long Baseline Array (VLBA) and the Global Millimeter VLBI Array (GMVA).
The simultaneity of the observations resulted in an unique opportunity to study the core shift of \C\ \citep{Rybicki79}, and by extrapolation pinpoint the jet apex.
For this study we used the already published images in \citep{Kim19}.
In order to determine the core shift, we calculated the the distance of the intensity peaks of the core region, at 15 and 43\,GHz, relative to the 86\,GHz intensity peak (which we placed at the origin).
This calculation was done by employing the standard two dimensional cross-correlation and determining where the cross-correlation coefficient peaked, for the 15-43 and 43-86\,GHz pairs.
For the alignment of the images, we chose optically thin, clearly defined regions, to perform the cross-correlation.
Specifically, for the 15-43\,GHz pair we used the C3 region, and for the 43-86\,GHz pair the region $\sim$1\,mas south of the 86\,GHz VLBI core.
The distances as a function of the frequency are plotted in Fig. \ref{fig:CS}.

%through back-extrapolation.

Following \cite{Lobanov98a}, we fitted a power law of the form:
\begin{equation}
    \Delta r_{\text {core }}=r_{0}\left(\left(\frac{v}{86 \mathrm{GHz}}\right)^{-1 / k_{r}}-1\right)[\mu \mathrm{as}], \label{Eq:CS}
\end{equation}
where  $k_r = [2n + b(3 - 2\alpha_{\rm thin} ) - 2]/(5 - 2\alpha_{\rm thin} )$, $n$ and $b$ are the particle number density and magnetic field strength power-law indices, respectively, and $\alpha_{\rm thin}$ is the optically thin spectral index. 
With the physically motivated equipartition assumption that $k_r = 1$ \cite{Lobanov98a, Hada11, Pushkarev12, Fromm13}, we find that the jet apex, and by extension the black hole of \C\ is at a distance of $83\pm7$\,$\mu$as upstream of the 86\,GHz VLBI core (see Fig. \ref{fig:CS}), in good agreement with \cite{Oh22}.
Our result, which is further supported by the position angle of the counter-jet reported at larger distances (see \citep{Vermeulen94}), points to a different scenario than the one presented in \citep{Giovannini18}.
In that work, the authors estimate a core shift of 0.03\,mas from the 22\,GHz VLBI core, under the assumption that the emission upstream of the core is the sub-mas counter-jet.

\begin{figure}
    \centering
    \includegraphics[width=1\columnwidth]{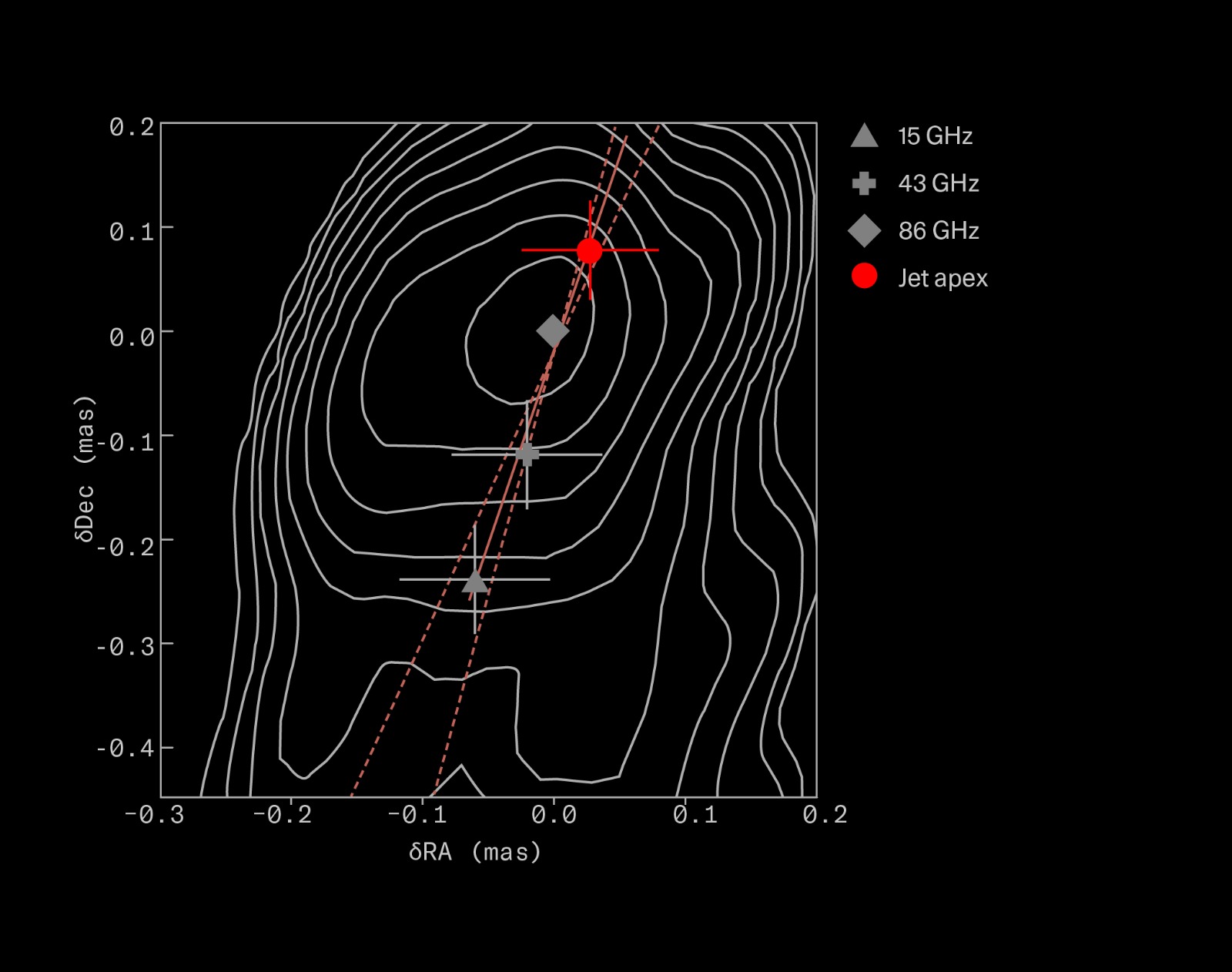}
    \caption{\small \sl Intensity peaks at 15, 43, 86\,GHz and extrapolated jet apex. The intensity peaks are denoted with the grey symbols, the jet apex with the red dot. They are super-imposed on 86\,GHz image contours, which start at 0.1\% of the image peak (1.82\,Jy/beam) and then increase in steps of two. The solid red line marks the extrapolated jet flow direction and the dashed red lines outline the 99\% confidence interval of the fit.}
    \label{fig:Position}
\end{figure}

In Fig. \ref{fig:Position} we plot the position of the intensity peaks of the core region at the three observed frequencies over image contours of the 86\,GHz observations. 
The jet apex appears upstream, north-west of the 86\,GHz intensity peak, at a position angle (P. A.) of -20$^\circ$\,$\pm$\,14$^\circ$.
Assuming a viewing angle in the $20^\circ-65^\circ$ range \citep{Abdo09, Fujita17}, the  de-projected distance of the jet apex to the 86\,GHz intensity peak is $0.028-0.11$\,pc, or $400-1500$\,R$_s$ (assuming a black hole mass of $9\times10^8\,M_\odot$ \citep{Scharwaechter13}). 

\section{Spectral index}
\noindent
A byproduct of the cross-correlation analysis are spectral index maps.
Here we define the spectral index as S$\,\propto\nu^{+\alpha}$. For the analysis we focus
on the two adjacent frequency pairs, because the large difference in beam size between 
15 and 86\,GHz causes beam blending effects. 

\begin{figure}
    \centering
    \includegraphics[width=1\columnwidth]{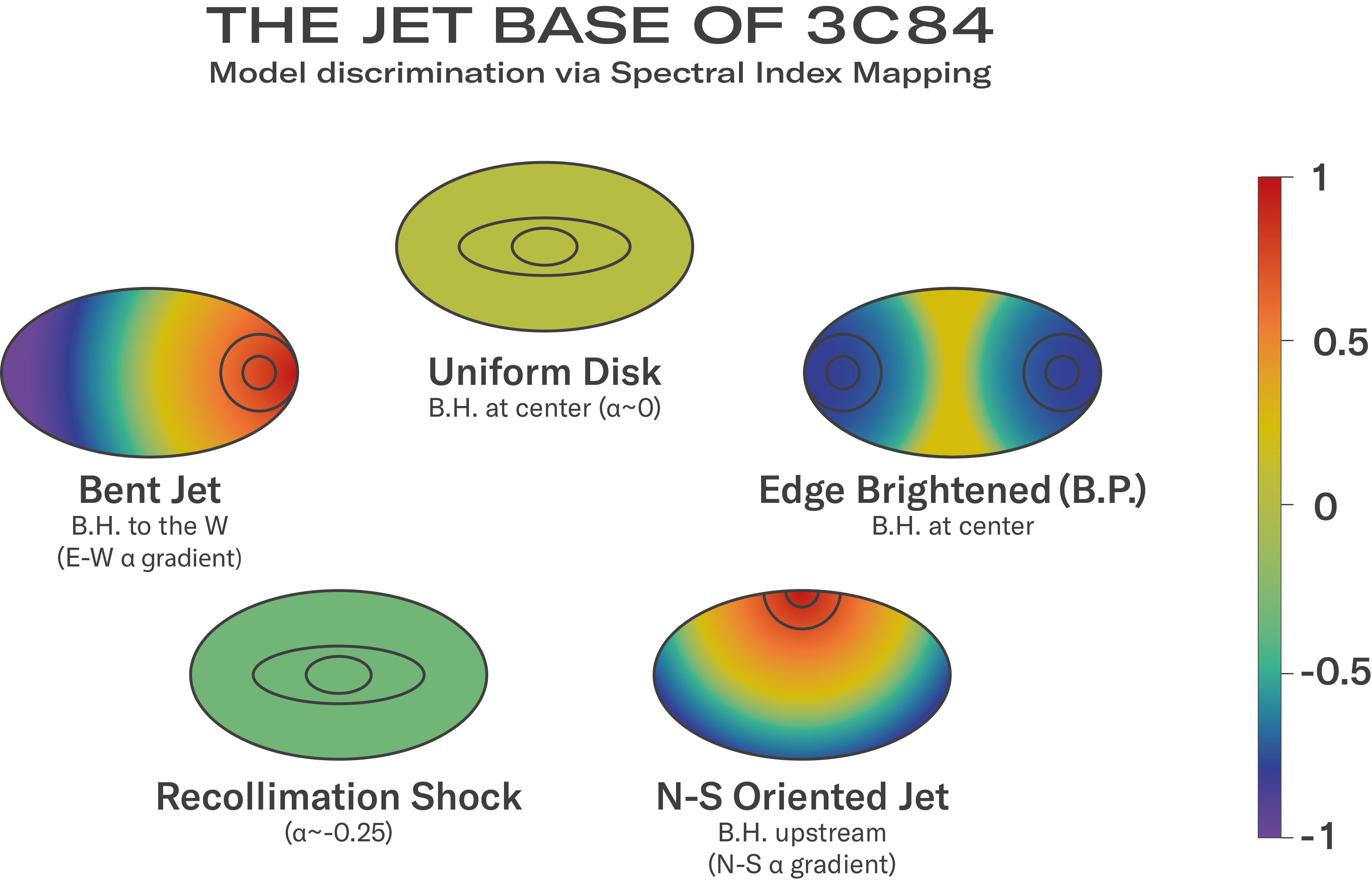}
    \caption{\small \sl Illustration of the expected spectral index, depending on the location and true nature of the VLBI core of \C.}
    \label{fig:SpI}
\end{figure}

Figure \ref{fig:SpI} illustrates different scenarios for the appearance of spectral index maps of the nuclear region, depending on the underlying physical model.
Starting from the top and in clockwise direction:
(i) for a black hole at the center surrounded by a uniform disk one may expect
%if the black hole were at the centre of the nucleus, surrounded by a uniform disk, we would expect to see 
a relatively uniform, flattish spectral index distribution. 
(ii) A Blandford \& Payne powered jet \cite{Blandford82} would lead to an edge brightened jet, manifesting in a steep spectral index east and west of the black hole, and flat at the center. 
(iii) An upstream located black hole would produce a smooth gradient in the north-south direction. 
(iv) An identification of the VLBI core as a recollimation shock (at some arbitrary distance from the BH) would lead to a flat to moderately steep spectral index.
(v) Finally, if the black hole is located at the western edge of the jet (c.f. \cite{Punsly21}) and if the inner (sub-mas) jet is strongly misaligned with the mas-scale jet, we would expect a spectral index gradient in the east-west direction.

\begin{figure}
    \centering
    \includegraphics[width=1\columnwidth]{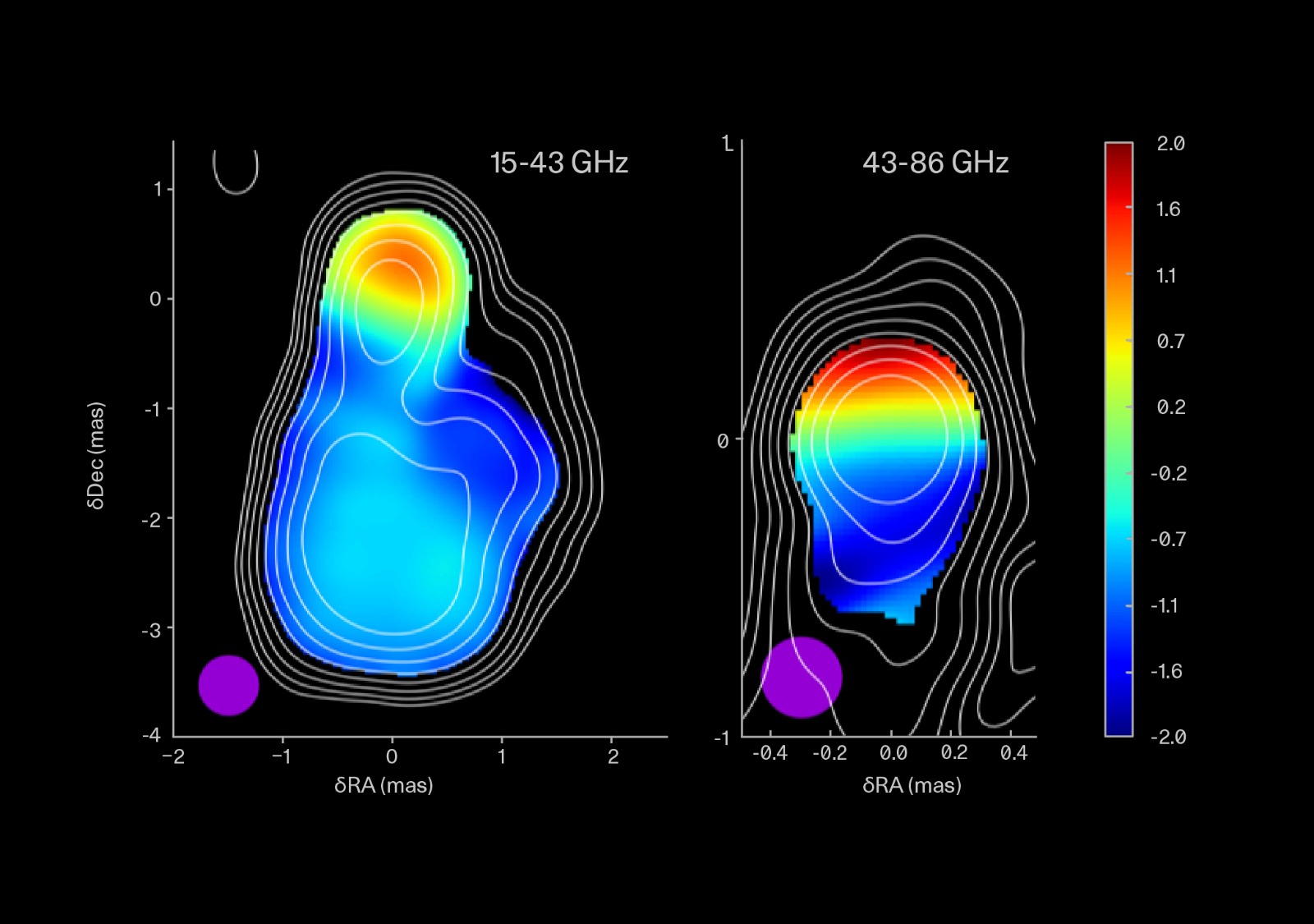}
    \caption{\small \sl Spectral index maps of \C\ at 15-43\,GHz and 43-86\,GHz. A prominent spectral index in the north-south direction is revealed. Left: 15-43\, GHz spectral index map, super-imposed on the 15\,GHz total intensity contours. Right: 43-86\, GHz spectral index map, super-imposed on the 43\,GHz total intensity contours.}
    \label{fig:SpI_obs}
\end{figure}

In Fig. \ref{fig:SpI_obs} we present our spectral index analysis. 
The presence of a north-south oriented spectral index gradient is obvious from both frequency pairs. We note the measurement of the spectral index gradient at 43-86\,GHz 
for the first time and at such high angular resolution. Comparing our results to the illustrations of Fig. \ref{fig:SpI}, they appear consistent with a jet apex location north of the 86\,GHz VLBI core and consequently a black hole location upstream of the visible portion of the jet.

\section{Magnetic field}
\noindent
Utilising a number of assumptions, namely that the jet is described by the 
Blandford \& Königl model \citep{Blandford79}, and is synchrotron self-absorbed, we can compute the magnetic field, as described in \citep{Lobanov98a, Fromm13}.
At the extrapolated jet apex location, we estimate the magnetic field to be $\sim$\,$2-4$\,G (see \cite{Paraschos21} for more details).
We can also compare our result to other prominent AGN, if we extrapolate the magnetic field strength
closer to the BH location (cf. \citep{Paraschos21}), for example to $\simeq$10\,R$_s$. At this distance the magnetic field $B_{10\,R_s}$ is $70-600$\,G, which compares well to M87 and NGC\,1052 \citep{Kim18, Baczko16}.
We note that a lower frequencies (15\,GHz),
a total of $\sim$\,100 quasar and BL Lac jets were analysed by \citep{Pushkarev12}. Based on their core shifts, an average magnetic field strength of $\sim$ $400-900$\,mG was obtained, at a distance of 1\,pc from the jet apex.
In \C, the magnetic field at the same distance from the core amounts to $60-180$\,mG, which is $4-6$ times lower
than the typical magnetic fields of the AGN studied by \cite{Pushkarev12}.

Besides the magnetic field strength, we also may set some constraints on its configuration. Our analysis points to a mixed toroidal/poloidal configuration ($b\approx1.7$), with the assumption that the total number of particles passing through each cross-section is conserved, and using the result by \cite{Giovannini18} that the jet radius $R$ and the distance to the core $z$ are connected by a power-law of the form $z\propto R^{0.21}$.

The visual appearance of an edge-brightened inner jet suggests transverse jet stratification, which is a natural consequence, for example from a combination of the Blandford \& Payne and a Blandford \& Znajek \cite{Blandford77} jet launching scenarios.
Overall, the magnetic field appears strong near the BH, compared to the upper limit estimation (see Eq. 8.35 in \citep{Ghisellini13}) from the total jet power \cite{Abdo09, Magic18}, in support of magnetic jet launching \citep{Narayan03, Tchkhovskoy11}.

\section{Conclusions}
\noindent

\begin{enumerate}
    \item The black hole in \C\ is located $\geq$ 400-1500\,R$_s$ upstream of the 86\,GHz VLBI core.
    \item A strong spectral index gradient in north-south direction is discovered, rendering \C\ a suitable target for core-shift analysis including the highest VLBI frequencies.  
      
    \item The magnetic field at 10\,R$_s$ is computed to be 70-600\,G, which compares well with other nearby radio galaxies, such as M\,87 or NGC\,1052.
    \item At distances comparable to the jet apex, we found a mixed poloidal-toroidal magnetic field configuration, pointing perhaps to a stratified combination of Blandford \& Payne and Blandford \& Znajek jet launching mechanism.
\end{enumerate}

The above analysis and results are described in more detail in \citep{Paraschos21}.

\section{Outlook}
\noindent
Although our results put new limits on the location of the jet apex and the BH in \C, the opening angle of the inner jet and the polarisation at the jet base are not yet
unambiguously determined. The accurate determination of the inner jet velocity and the direction of the component motion is yet another important topic. Further VLBI imaging observations with sufficient high angular resolution (GMVA at 43/86 GHz, EHT at 230/345 GHz) will address these topics. 
Specifically, already planned quasi-simultaneous observations at 22 and 43\,GHz with the Global-EVN  and 86 and 230\,GHz observations with the GMVA and EHT should yield better polarimetric data, providing a more detailed insight into the magnetic field properties for the inner jet region in \C.

\acknowledgments
\noindent
We thank T. Savolainen for providing software to calculate two dimensional cross-correlations.
This work makes use of 15, 37\,GHz and 230\,GHz light curves kindly provided by the Owen's Valley Radio Observatory, Metsähovi Radio Observatory and the Submillimeter Array, respectively.
G. F. Paraschos is supported for this research by the International Max-Planck Research School (IMPRS) for Astronomy and Astrophysics at the University of Bonn and Cologne. This research has made use of data obtained with the Global Millimeter VLBI Array (GMVA), which consists of telescopes operated by the MPIfR, IRAM, Onsala, Metsähovi, Yebes, the Korean VLBI Network, the Green Bank Observatory and the VLBA (NRAO). The Submillimeter Array (SMA) is a joint project between the Smithsonian Astrophysical Observatory and the Academia Sinica Institute of Astronomy and Astrophysics and is funded by the Smithsonian Institution and the Academia Sinica. This research has also made use of data from the Owen's Valley Radio Observatory 40-m monitoring program \citep{Richards11}, supported by private funding from the California Insitute of Technology and the Max Planck Institute for Radio Astronomy, and by NASA grants NNX08AW31G, NNX11A043G, and NNX14AQ89G and NSF grants AST-0808050 and AST-1109911.
Finally, this research makes use the publicly available $\gamma$-ray light curve of NGC\,1275 (\url{https://fermi.gsfc.nasa.gov/ssc/data/access/lat/msl_lc/source/NGC_1275}).

% Bibliography
\bibliographystyle{JHEP}
\bibliography{references}

\providecommand{\href}[2]{#2}\begingroup\raggedright\begin{thebibliography}{10}

\bibitem{Baade54}
W.~{Baade} and R.~{Minkowski}, \emph{{On the Identification of Radio
  Sources.}}, \href{https://doi.org/10.1086/145813}{\emph{\apj} {\bfseries 119}
  (1954) 215}.

\bibitem{Walker00}
R.~C. {Walker}, V.~{Dhawan}, J.~D. {Romney}, K.~I. {Kellermann} and R.~C.
  {Vermeulen}, \emph{{VLBA Absorption Imaging of Ionized Gas Associated with
  the Accretion Disk in NGC 1275}},
  \href{https://doi.org/10.1086/308372}{\emph{\apj} {\bfseries 530} (2000) 233}
  [\href{https://arxiv.org/abs/astro-ph/9909365}{{\ttfamily
  astro-ph/9909365}}].

\bibitem{Dhawan90}
V.~{Dhawan}, N.~{Bartel}, A.~E.~E. {Rogers}, T.~P. {Krichbaum}, A.~{Witzel},
  D.~A. {Graham} et~al., \emph{{Further 7 Millimeter VLBI Observations of 3C 84
  and Other Sources with 100 Microarcsecond Angular Resolution}},
  \href{https://doi.org/10.1086/185808}{\emph{\apjl} {\bfseries 360} (1990)
  L43}.

\bibitem{Krichbaum92}
T.~P. {Krichbaum}, A.~{Witzel}, D.~A. {Graham}, W.~{Alef}, I.~I.~K.
  {Pauliny-Toth}, C.~A. {Hummel} et~al., \emph{{The evolution of the sub-parsec
  structure of 3C 84 at 43 GHz.}}, {\emph{\aap} {\bfseries 260} (1992) 33}.

\bibitem{Nagai14}
H.~{Nagai}, T.~{Haga}, G.~{Giovannini}, A.~{Doi}, M.~{Orienti}, F.~{D'Ammando}
  et~al., \emph{{Limb-brightened Jet of 3C 84 Revealed by the 43 GHz
  Very-Long-Baseline-Array Observation}},
  \href{https://doi.org/10.1088/0004-637X/785/1/53}{\emph{\apj} {\bfseries 785}
  (2014) 53} [\href{https://arxiv.org/abs/1402.5930}{{\ttfamily 1402.5930}}].

\bibitem{Hodgson21}
J.~A. {Hodgson}, B.~{Rani}, J.~{Oh}, A.~{Marscher}, S.~{Jorstad}, Y.~{Mizuno}
  et~al., \emph{{A Detailed Kinematic Study of 3C 84 and Its Connection to
  {\ensuremath{\gamma}}-Rays}},
  \href{https://doi.org/10.3847/1538-4357/abf6dd}{\emph{\apj} {\bfseries 914}
  (2021) 43} [\href{https://arxiv.org/abs/2104.03081}{{\ttfamily 2104.03081}}].

\bibitem{Dent66}
W.~A. {Dent}, \emph{{Variation in the Radio Emission from the Seyfert Galaxy
  NGC 1275}}, \href{https://doi.org/10.1086/148674}{\emph{\apj} {\bfseries 144}
  (1966) 843}.

\bibitem{Vermeulen94}
R.~C. {Vermeulen}, A.~C.~S. {Readhead} and D.~C. {Backer}, \emph{{Discovery of
  a Nuclear Counterjet in NGC 1275: A New Way to Probe the Parsec-Scale
  Environment}}, \href{https://doi.org/10.1086/187433}{\emph{\apjl} {\bfseries
  430} (1994) L41}.

\bibitem{Walker94}
R.~C. {Walker}, J.~D. {Romney} and J.~M. {Benson}, \emph{{Detection of a VLBI
  Counterjet in NGC 1275: A Possible Probe of the Parsec-Scale Accretion
  Region}}, \href{https://doi.org/10.1086/187434}{\emph{\apjl} {\bfseries 430}
  (1994) L45}.

\bibitem{Fujita17}
Y.~{Fujita} and H.~{Nagai}, \emph{{Discovery of a new subparsec counterjet in
  NGC 1275: the inclination angle and the environment}},
  \href{https://doi.org/10.1093/mnrasl/slw217}{\emph{\mnras} {\bfseries 465}
  (2017) L94} [\href{https://arxiv.org/abs/1609.04017}{{\ttfamily
  1609.04017}}].

\bibitem{Wajima20}
K.~{Wajima}, M.~{Kino} and N.~{Kawakatu}, \emph{{Constraints on the
  Circumnuclear Disk through Free-Free Absorption in the Nucleus of 3C 84 with
  KaVA and KVN at 43 and 86 GHz}},
  \href{https://doi.org/10.3847/1538-4357/ab88a0}{\emph{\apj} {\bfseries 895}
  (2020) 35} [\href{https://arxiv.org/abs/2004.06589}{{\ttfamily 2004.06589}}].

\bibitem{Kim19}
J.~Y. {Kim}, T.~P. {Krichbaum}, A.~P. {Marscher}, S.~G. {Jorstad}, I.~{Agudo},
  C.~{Thum} et~al., \emph{{Spatially resolved origin of millimeter-wave linear
  polarization in the nuclear region of 3C 84}},
  \href{https://doi.org/10.1051/0004-6361/201832920}{\emph{\aap} {\bfseries
  622} (2019) A196} [\href{https://arxiv.org/abs/1811.07815}{{\ttfamily
  1811.07815}}].

\bibitem{Dhawan98}
V.~{Dhawan}, K.~I. {Kellermann} and J.~D. {Romney}, \emph{{Kinematics of the
  Nucleus of NGC 1275 (3C 84)}},
  \href{https://doi.org/10.1086/311313}{\emph{\apjl} {\bfseries 498} (1998)
  L111}.

\bibitem{Britzen19}
S.~{Britzen}, C.~{Fendt}, M.~{Zaja{\v{c}}ek}, F.~{Jaron}, I.~{Pashchenko},
  M.~F. {Aller} et~al., \emph{{3C 84: Observational Evidence for Precession and
  a Possible Relation to TeV Emission}},
  \href{https://doi.org/10.3390/galaxies7030072}{\emph{Galaxies} {\bfseries 7}
  (2019) 72}.

\bibitem{Linhoff21}
L.~{Linhoff}, A.~{Sandrock}, M.~{Kadler}, D.~{Els{\"a}sser} and W.~{Rhode},
  \emph{{Excluding possible sites of high-energy emission in 3C 84}},
  \href{https://doi.org/10.1093/mnras/staa3521}{\emph{\mnras} {\bfseries 500}
  (2021) 4671}.

\bibitem{Richards11}
J.~L. {Richards}, W.~{Max-Moerbeck}, V.~{Pavlidou}, O.~G. {King}, T.~J.
  {Pearson}, A.~C.~S. {Readhead} et~al., \emph{{Blazars in the Fermi Era: The
  OVRO 40 m Telescope Monitoring Program}},
  \href{https://doi.org/10.1088/0067-0049/194/2/29}{\emph{\apjs} {\bfseries
  194} (2011) 29} [\href{https://arxiv.org/abs/1011.3111}{{\ttfamily
  1011.3111}}].

\bibitem{Kocevski21}
{Fermi Large Area Telescope Collaboration}, \emph{{10-year Fermi LAT point
  source catalog}}, {\emph{The Astronomer's Telegram} {\bfseries 15110} (2021)
  1}.

\bibitem{Rybicki79}
G.~B. {Rybicki} and A.~P. {Lightman}, \emph{{Radiative processes in
  astrophysics}}. 1979.

\bibitem{Lobanov98a}
A.~P. {Lobanov}, \emph{{Spectral distributions in compact radio sources. I.
  Imaging with VLBI data}},
  \href{https://doi.org/10.1051/aas:1998446}{\emph{\aaps} {\bfseries 132}
  (1998) 261} [\href{https://arxiv.org/abs/astro-ph/9804112}{{\ttfamily
  astro-ph/9804112}}].

\bibitem{Hada11}
K.~{Hada}, A.~{Doi}, M.~{Kino}, H.~{Nagai}, Y.~{Hagiwara} and N.~{Kawaguchi},
  \emph{{An origin of the radio jet in M87 at the location of the central black
  hole}}, \href{https://doi.org/10.1038/nature10387}{\emph{\nat} {\bfseries
  477} (2011) 185}.

\bibitem{Pushkarev12}
A.~B. {Pushkarev}, T.~{Hovatta}, Y.~Y. {Kovalev}, M.~L. {Lister}, A.~P.
  {Lobanov}, T.~{Savolainen} et~al., \emph{{MOJAVE: Monitoring of Jets in
  Active galactic nuclei with VLBA Experiments. IX. Nuclear opacity}},
  \href{https://doi.org/10.1051/0004-6361/201219173}{\emph{\aap} {\bfseries
  545} (2012) A113} [\href{https://arxiv.org/abs/1207.5457}{{\ttfamily
  1207.5457}}].

\bibitem{Fromm13}
C.~M. {Fromm}, E.~{Ros}, M.~{Perucho}, T.~{Savolainen}, P.~{Mimica},
  M.~{Kadler} et~al., \emph{{Catching the radio flare in CTA 102. III.
  Core-shift and spectral analysis}},
  \href{https://doi.org/10.1051/0004-6361/201321784}{\emph{\aap} {\bfseries
  557} (2013) A105} [\href{https://arxiv.org/abs/1306.6208}{{\ttfamily
  1306.6208}}].

\bibitem{Oh22}
J.~{Oh}, J.~A. {Hodgson}, S.~{Trippe}, T.~P. {Krichbaum}, M.~{Kam}, G.~F.
  {Paraschos} et~al., \emph{{A persistent double nuclear structure in 3C 84}},
  \href{https://doi.org/10.1093/mnras/stab3056}{\emph{\mnras} {\bfseries 509}
  (2022) 1024} [\href{https://arxiv.org/abs/2110.09811}{{\ttfamily
  2110.09811}}].

\bibitem{Giovannini18}
G.~{Giovannini}, T.~{Savolainen}, M.~{Orienti}, M.~{Nakamura}, H.~{Nagai},
  M.~{Kino} et~al., \emph{{A wide and collimated radio jet in 3C84 on the scale
  of a few hundred gravitational radii}},
  \href{https://doi.org/10.1038/s41550-018-0431-2}{\emph{Nature Astronomy}
  {\bfseries 2} (2018) 472} [\href{https://arxiv.org/abs/1804.02198}{{\ttfamily
  1804.02198}}].

\bibitem{Abdo09}
A.~A. {Abdo}, M.~{Ackermann}, M.~{Ajello}, K.~{Asano}, L.~{Baldini},
  J.~{Ballet} et~al., \emph{{Fermi Discovery of Gamma-ray Emission from NGC
  1275}}, \href{https://doi.org/10.1088/0004-637X/699/1/31}{\emph{\apj}
  {\bfseries 699} (2009) 31} [\href{https://arxiv.org/abs/0904.1904}{{\ttfamily
  0904.1904}}].

\bibitem{Scharwaechter13}
J.~{Scharw{\"a}chter}, P.~J. {McGregor}, M.~A. {Dopita} and T.~L. {Beck},
  \emph{{Kinematics and excitation of the molecular hydrogen accretion disc in
  NGC 1275}}, \href{https://doi.org/10.1093/mnras/sts502}{\emph{\mnras}
  {\bfseries 429} (2013) 2315}
  [\href{https://arxiv.org/abs/1211.6750}{{\ttfamily 1211.6750}}].

\bibitem{Blandford82}
R.~D. {Blandford} and D.~G. {Payne}, \emph{{Hydromagnetic flows from accretion
  disks and the production of radio jets.}},
  \href{https://doi.org/10.1093/mnras/199.4.883}{\emph{\mnras} {\bfseries 199}
  (1982) 883}.

\bibitem{Punsly21}
B.~{Punsly}, H.~{Nagai}, T.~{Savolainen} and M.~{Orienti}, \emph{{Observing the
  Time Evolution of the Multicomponent Nucleus of 3C 84}},
  \href{https://doi.org/10.3847/1538-4357/abe69f}{\emph{\apj} {\bfseries 911}
  (2021) 19} [\href{https://arxiv.org/abs/2102.07272}{{\ttfamily 2102.07272}}].

\bibitem{Blandford79}
R.~D. {Blandford} and A.~{K{\"o}nigl}, \emph{{Relativistic jets as compact
  radio sources.}}, \href{https://doi.org/10.1086/157262}{\emph{\apj}
  {\bfseries 232} (1979) 34}.

\bibitem{Paraschos21}
G.~F. {Paraschos}, J.~Y. {Kim}, T.~P. {Krichbaum} and J.~A. {Zensus},
  \emph{{Pinpointing the jet apex of 3C 84}},
  \href{https://doi.org/10.1051/0004-6361/202140776}{\emph{\aap} {\bfseries
  650} (2021) L18} [\href{https://arxiv.org/abs/2106.04918}{{\ttfamily
  2106.04918}}].

\bibitem{Kim18}
J.~Y. {Kim}, T.~P. {Krichbaum}, R.~S. {Lu}, E.~{Ros}, U.~{Bach}, M.~{Bremer}
  et~al., \emph{{The limb-brightened jet of M87 down to the 7 Schwarzschild
  radii scale}}, \href{https://doi.org/10.1051/0004-6361/201832921}{\emph{\aap}
  {\bfseries 616} (2018) A188}
  [\href{https://arxiv.org/abs/1805.02478}{{\ttfamily 1805.02478}}].

\bibitem{Baczko16}
A.~K. {Baczko}, R.~{Schulz}, M.~{Kadler}, E.~{Ros}, M.~{Perucho}, T.~P.
  {Krichbaum} et~al., \emph{{A highly magnetized twin-jet base pinpoints a
  supermassive black hole}},
  \href{https://doi.org/10.1051/0004-6361/201527951}{\emph{\aap} {\bfseries
  593} (2016) A47} [\href{https://arxiv.org/abs/1605.07100}{{\ttfamily
  1605.07100}}].

\bibitem{Blandford77}
R.~D. {Blandford} and R.~L. {Znajek}, \emph{{Electromagnetic extraction of
  energy from Kerr black holes.}},
  \href{https://doi.org/10.1093/mnras/179.3.433}{\emph{\mnras} {\bfseries 179}
  (1977) 433}.

\bibitem{Ghisellini13}
G.~{Ghisellini}, \emph{{Radiative Processes in High Energy Astrophysics}},
  vol.~873. 2013,
  \href{https://doi.org/10.1007/978-3-319-00612-3}{10.1007/978-3-319-00612-3}.

\bibitem{Magic18}
{MAGIC Collaboration}, S.~{Ansoldi}, L.~A. {Antonelli}, C.~{Arcaro},
  D.~{Baack}, A.~{Babi{\'c}} et~al., \emph{{Gamma-ray flaring activity of
  NGC1275 in 2016-2017 measured by MAGIC}},
  \href{https://doi.org/10.1051/0004-6361/201832895}{\emph{\aap} {\bfseries
  617} (2018) A91} [\href{https://arxiv.org/abs/1806.01559}{{\ttfamily
  1806.01559}}].

\bibitem{Narayan03}
R.~{Narayan}, I.~V. {Igumenshchev} and M.~A. {Abramowicz}, \emph{{Magnetically
  Arrested Disk: an Energetically Efficient Accretion Flow}},
  \href{https://doi.org/10.1093/pasj/55.6.L69}{\emph{\pasj} {\bfseries 55}
  (2003) L69} [\href{https://arxiv.org/abs/astro-ph/0305029}{{\ttfamily
  astro-ph/0305029}}].

\bibitem{Tchkhovskoy11}
A.~{Tchekhovskoy}, R.~{Narayan} and J.~C. {McKinney}, \emph{{Efficient
  generation of jets from magnetically arrested accretion on a rapidly spinning
  black hole}},
  \href{https://doi.org/10.1111/j.1745-3933.2011.01147.x}{\emph{\mnras}
  {\bfseries 418} (2011) L79}
  [\href{https://arxiv.org/abs/1108.0412}{{\ttfamily 1108.0412}}].

\end{thebibliography}\endgroup
\end{document}